\documentclass[a4paper,12pt]{article}
\usepackage[utf8]{inputenc}

\usepackage{amsmath,amsfonts,amssymb}
\usepackage{bm}
\usepackage{physics}
\usepackage{tensor}
\usepackage{ascmac}
\usepackage{here}
\usepackage{xcolor}

\usepackage{graphicx}
\usepackage{bmpsize}
\graphicspath{{image/}}

\usepackage{mathrsfs}
\usepackage{authblk}
\usepackage{title}
\numberwithin{equation}{section}
\definecolor{DarkBlue}{rgb}{0,0,0.9} 

\usepackage{makecell}

\usepackage[top=30truemm,bottom=25truemm,left=25truemm,right=30truemm]{geometry}
\usepackage[colorlinks=true
,urlcolor=blue
,anchorcolor=blue
,citecolor=blue
,filecolor=blue
,linkcolor=blue
,menucolor=blue
,pagecolor=blue
,linktocpage=true
,pdfproducer=medialab
,pdfa=true
]{hyperref}

\usepackage{cite}

\usepackage[top=30truemm,bottom=25truemm,left=25truemm,right=30truemm]{geometry}
\usepackage[colorlinks=true
,urlcolor=blue
,anchorcolor=blue
,citecolor=blue
,filecolor=blue
,linkcolor=blue
,menucolor=blue
,pagecolor=blue
,linktocpage=true
,pdfproducer=medialab
,pdfa=true
]{hyperref}

\begin{document}

\title{Dynamical Formation of Self-Similar Wormholes}

\author[$a$,$b$]{\textbf{Yasutaka Koga}\thanks{\href{mailto:yasutaka.koga@oit.ac.jp}{yasutaka.koga@oit.ac.jp}}}
\author[$b$]{\textbf{Ryota Maeda} \thanks{\href{mailto:ryota.maeda@yukawa.kyoto-u.ac.jp}{ryota.maeda@yukawa.kyoto-u.ac.jp}}}
\author[$c$]{\textbf{Daiki 
Saito}
\thanks{\href{mailto:saito@tap.scphys.kyoto-u.ac.jp}{saito@tap.scphys.kyoto-u.ac.jp}}}

\author[$d$]{\textbf{Daisuke Yoshida}
\thanks{\href{mailto:dyoshida@math.nagoya-u.ac.jp}{dyoshida@math.nagoya-u.ac.jp}}}

\affil[$a$]{{\small	 Department of Information and Computer Science, Osaka Institute of Technology, Hirakata 573-0196, Japan}}
\affil[$b$]{{\small	Center for Gravitational Physics and Quantum Information, Yukawa Institute for Theoretical Physics, Kyoto University,\\ Kyoto 606-8502, Japan}}
\affil[$c$]{{\small	 Department of Physics, Kyoto University, Kyoto 606-8502, Japan}}
\affil[$d$]{{\small	 Department of Mathematics, Nagoya University, Nagoya 464-8602, Japan}}

\begin{flushright}
YITP-26-15, KUNS-3091
\end{flushright}

\maketitle
\thispagestyle{empty}

\begin{abstract}
We study spherically symmetric, self-similar wormhole solutions supported by colliding streams of negative-energy null dust, and their dynamical formation.
Under the assumption of self-similarity, the Einstein equations reduce to a system of ordinary differential equations, which we solve numerically under boundary conditions enforcing the existence of a minimal areal radius (the throat) on constant-time hypersurfaces.
For a sufficiently large throat radius, the resulting geometries remain regular at both spatial and future null infinity, while a singularity is retained in the past direction.
We then construct a dynamical formation scenario by patching together three regions: a Schwarzschild black hole, negative-energy Vaidya spacetimes, and the self-similar wormhole geometry.
These regions are joined across null shells using the Barrabès–Israel formalism, which provides explicit relations among the throat radius, the black hole's mass and the energy injection by the shell, demonstrating that an initial black hole can evolve into a wormhole.
Our analysis generalizes the formation model for static wormhole solutions proposed by Hayward and Koyama in 2004 to non-static wormhole solutions, offering a novel perspective on the formation of regular traversable wormholes.
\end{abstract}

\newpage
\setcounter{page}{1}

\tableofcontents

\section{Introduction}
Traversable wormholes, hypothetical spacetime configurations that allow causal communication between distant regions, have been investigated extensively since the pioneering works of Morris and Thorne~\cite{Morris:1988cz, PhysRevLett.61.1446}. 
Today, they serve as theoretical laboratories for exploring the interplay among geometry, exotic matter sources, and quantum effects, beyond their speculative use as transport devices for interstellar travel.
It has been pointed out~\cite{Morris:1988cz,Friedman:1993ty,Visser:1995cc,Hochberg:1997wp,Visser:2003yf,Friedman:2006tea} that, static wormhole solutions in general relativity necessitate the violation of the (averaged) null energy conditions~\cite{Ford:1994bj,Graham:2007va}.
This fact has motivated many studies of possible exotic fields that could support such geometries. For example, wormhole solutions with a phantom scalar field have been investigated in Refs.\cite{Bronnikov:1973fh,Ellis:1973yv,Armendariz-Picon:2002gjc,Makita:2025bao}.
Realization of the violation of the energy condition without assuming an exotic field has been studied, for example, considering the squeezed states of matter~\cite{Hochberg:1991tz}, holographic setup~\cite{Gao:2016bin,Maldacena:2017axo,Caceres:2018ehr,Maldacena:2018lmt,Kawamoto:2025oko} the Casimir energy of fields~\cite{Maldacena:2018gjk,Garattini:2019ivd}, and the higher derivative corrections to the Einstein equations~\cite{Kanai:2025zgq}.

Alongside the question of matter sources, considerable attention has been devoted to purely geometric constructions of wormhole spacetimes with minimal assumptions on the matter content.
For example, in Ref.~\cite{Hayward:2002pm}, Hayward constructed the analytic solution of a static and spherically symmetric wormhole.
In that work, the author just assumed that the energy-momentum tensor takes the form of the colliding null dust with the negative energy, without any further assumptions on the matter field. 
Due to this minimal assumption, the solution is locally applicable to any matter source that shares the same form of the energy-momentum tensor around the throat.
However, it cannot be regarded as a globally viable wormhole, since singularities appear in the spatial and null directions, as pointed out in Ref.~\cite{Koga:2025bqw}.
In Ref.~\cite{Koga:2025bqw}, the Hayward solution was extended to include the electric charges, and it is found that the singularity persists even in the presence of charges.
This indicates that singularities are a generic feature of static models with this type of energy-momentum tensor.

A different but related question concerns the dynamical formation of wormholes.
While black hole formation is widely regarded as a natural outcome of stellar collapse, the dynamical formation of wormholes from realistic initial conditions presents distinct theoretical challenges.
Several mechanisms have been proposed, such as irradiation of negative energy into black holes~\cite{Hayward:1998pp} and geometric patching that yields Roberts-type solutions as the final state~\cite{Maeda:2008bh}.
Among these studies,Hayward and Koyama~\cite{Hayward:2004wm,Koyama:2004uh} showed that injecting negative-energy flux into a Schwarzschild black hole can generate a static Hayward wormhole~\cite{Hayward:2002pm} through thin-shell matching based on the Barrabès–Israel formalism~\cite{Barrabes:1991ng}.
However, this construction inherits the same limitation as the Hayward solution~\cite{Hayward:2002pm}, namely the presence of spacetime singularities.

In this paper, we consider a non-static extension of Hayward’s model~\cite{Hayward:2002pm} and its application to a formation scenario~\cite{Hayward:2004wm,Koyama:2004uh}.
In particular, we impose self-similar symmetry to analyze time-dependent wormhole solutions in a simplified setting and investigate how the geometry—including the occurrence of singularities—evolves.
This work also aims to construct a comprehensive dynamical formation model by employing the Barrabès–Israel formalism~\cite{Barrabes:1991ng} with Schwarzschild black holes being initial conditions of formation of wormhole geometry.

The paper is organized as follows.
Section~\ref{SSWS} develops the self-similar wormhole solution.
Section~\ref{Formation} details the geometric matching for dynamical formation.
Section~\ref{Section} is dedicated to the summary and discussion.

\section{Self-similar wormhole solution}
\label{SSWS}

\subsection{Spherically symmetric spacetime with colliding null dust}
\label{SSCND}

We begin by introducing the spherically symmetric ansatz describing two
counter-propagating streams of null dust,
extending the static setup introduced by Hayward~\cite{Hayward:2002pm}. 
This matter model provides a particularly transparent way to realize violations of the null energy condition under spherical symmetry.
It therefore serves as a convenient starting point for constructing time-dependent wormhole geometries.

Let us adopt the generic spherically symmetric metric ansatz
\begin{align}
 \boldsymbol{g} = - \frac{\mathrm{e}^{B(u,v)}}{2
 \sqrt{A(u,v)}} (\boldsymbol{d}u \otimes \boldsymbol{d}v + \boldsymbol{d}v \otimes \boldsymbol{d}u) + uvA(u,v) \boldsymbol{d}\Omega^2. \label{metric}
\end{align}
Here, $u$ and $v$ denote the null coordinates, and $\boldsymbol{d}\Omega^2$ denotes the metric of the unit two-sphere. 
We assume $\bm{\partial}_{u}$ and $\bm{\partial}_{v}$ point future direction without loss of generality.

The energy–momentum tensor is taken to be that of two null dust streams
propagating along $\partial_{u}$ and $\partial_{v}$ directions:
\begin{align}
 \boldsymbol{T} = T_{uu}(u,v) \boldsymbol{d}u \otimes \boldsymbol{d}u  + T_{vv}(u,v) \boldsymbol{d}v \otimes \boldsymbol{d}v.
\end{align}
Solving the conservation equations $\nabla_{\mu} T^{\mu}{}_{\nu} = 0$, we obtain
\begin{align}
 T_{uu}(u,v) = \frac{1}{8 \pi G} \frac{\tau_{u}(u)}{uv A(u,v)}, \qquad  T_{vv}(u,v) = \frac{1}{8 \pi G} \frac{\tau_{v}(v)}{u vA(u,v)}.
 \label{Tuuvv}
\end{align}
Here, we have introduced the arbitrary functions $\tau_{u}(u)$ and $\tau_{v}(v)$, which depend only on $u$ and $v$, respectively.

The Einstein Equation $G_{\mu\nu} = 8 \pi G T_{\mu\nu}$ reduces to two coupled wave equations:
\begin{align}
 \partial_{u} \partial_{v} \qty(uvA(u,v)) &= - \frac{\mathrm{e}^{B(u,v)}}{2\sqrt{A(u,v)}}, \label{waveeqY} \\
 \partial_{u} \partial_{v} B(u,v) &= \frac{1}{4} \frac{\mathrm{e}^{B(u,v)}}{uvA^{\frac{3}{2}}}, \label{waveeqX}
\end{align}
together with
\begin{align}
 \tau_{u}(u) = \qty(\frac{1}{2u}+ \partial_{u}B)\partial_{u}(uvA)-\partial^2_{u}(uvA), \label{cusol} \\
 \tau_{v}(v) = \qty(\frac{1}{2v}+ \partial_{v}B)\partial_{v}(uvA)-\partial^2_{v}(uvA).\label{cvsol}
\end{align}
The consistency conditions $\partial_{v} \tau_{u} = 0$ and $\partial_{u} \tau_{v} = 0$ follow automatically from Eqs.~\eqref{waveeqY} and \eqref{waveeqX}.
Thus, for any functions $A$ and $B$ satisfying the wave equations, the metric \eqref{metric} solves the Einstein equations with null dust, with $\tau_{u}$ and $\tau_{v}$ determined by Eqs.~\eqref{cusol} and \eqref{cvsol}.

By solving Eq.~\eqref{waveeqY} for $B$, we obtain
\begin{align}
 B(u,v) = \log \left( - 2\sqrt{A(u,v)} \partial_{u} \partial_{v} \qty(uvA) \right),
\end{align}
and substituting this into Eq.~\eqref{waveeqX} yields a closed equation for $A$,
\begin{align}
&uvA\qty[\qty(\partial_{u}\partial_{v}(uvA))^3-uvA\partial^2_{u}\partial_{v}(uvA)\partial_{u}\partial^2_{v}(uvA) + uvA\partial_{u}\partial_{v}(uvA)\partial^2_{u}\partial^2_{v}(uvA)] \nonumber \\
&-\frac{1}{2}\partial_{u}(uvA)\partial_{v}(uvA)\qty(\partial_{u}\partial_{v}(uvA))^2= 0. \label{eqY}
\end{align}
After eliminating $B$, the metric can be expressed as
\begin{align}
\boldsymbol{g} =  \partial_{u} \partial_{v} \left(u v A(u,v) \right) (\boldsymbol{d}u \otimes \boldsymbol{d}v + \boldsymbol{d}v \otimes \boldsymbol{d}u)
 + uvA(u,v) \boldsymbol{d}\Omega^2.
\end{align}
Hence, any solution of Eq.~\eqref{eqY} generates an exact spherically symmetric solution of the Einstein equations with colliding null dust.

\subsection{Equations of motion with self-similarity}

In the previous subsection, we discussed a general class of spherically symmetric solutions sourced by colliding null dust, which is parametrized by a function $A(u,v)$. 
However, the partial differential equation~\eqref{eqY} is highly nonlinear, making the analysis difficult.
To proceed, we now impose an additional simplifying symmetry: self-similarity.
Specifically, we assume the existence of a homothetic Killing vector
\begin{align}
 \boldsymbol{\xi} = \xi^{u}(u,v)  \boldsymbol{\partial}_{u} + \xi^{v}(u,v) \boldsymbol{\partial}_{v},
\end{align}
which satisfies the homothetic Killing equation
\begin{align}
 (\mathsterling_{\boldsymbol{\xi}} \boldsymbol{g})_{\mu\nu} = \nabla_{\mu} \xi_{\nu} + \nabla_{\nu} \xi_{\mu} = 2 g_{\mu\nu}.
\end{align}
We further assume that $\boldsymbol{\xi}$ is a future directed timelike vector.
From the equation, $\xi^{u}$ and $\xi^{v}$ must depend only on $u$ and $v$, respectively.
Since $\boldsymbol{\xi}$ is assumed to be timelike, both $\xi^{u}$ and $\xi^{v}$ are non-vanishing and, therefore, we can choose the coordinates so that
\begin{align}
 \boldsymbol{\xi} = u \boldsymbol{\partial}_{u} + v \boldsymbol{\partial}_{v}.
 \label{expression of xi}
\end{align}
We note that $\bm{\xi}$ is timelike and future directed only when $u > 0, v>0$ in these coordinates and we will focus on this case in the following.
We note that at this stage, we still have a degrees of freedom for the choice of $u,v$ coordinates up to rescaling.
Under this assumption, the remaining components of the homothetic Killing equation  are solved, showing that $A$ and $B$ depend only on the ratio of $u$ and $v$.

Let us introduce new coordinates $(t,l)$ by
\begin{align}
 u v = L^2 \mathrm{e}^{2 t} \qquad \frac{v}{u} = \mathrm{e}^{2 l},
\end{align} 
where $L$ is a constant parameter with the dimension of the length.
In these new coordinates, the homothetic Killing vector is expressed as
\begin{align}
 \boldsymbol{\xi} =  \boldsymbol{\partial}_{t},
\end{align}
which implies that the geometry is self-similar in the ``time'' direction.
Indeed, the metric takes the conformal static form 
\begin{align}
 \boldsymbol{g} = \mathrm{e}^{2 t}L^2 \qty[
\frac{\mathrm{e}^{B(l)}}{\sqrt{A(l)}} \left( - \boldsymbol{d}t  \otimes  \boldsymbol{d}t + \boldsymbol{d}l  \otimes  \boldsymbol{d}l \right)
+ A(l) \boldsymbol{d}\Omega^2], \label{mettl}
\end{align}
with the areal radius being written as $r=\mathrm{e}^{t}L\sqrt{A(l)}$.
In these coordinates, the coupled wave equations~\eqref{waveeqY} and \eqref{waveeqX} reduce to
\begin{align}
&A''(l) - 4 A(l) - 2  \frac{\mathrm{e}^{B(l)}}{\sqrt{A(l)}} = 0, \label{Aeq0} \\
&B''(l)  +  \frac{\mathrm{e}^{B(l)}}{A(l)^{3/2}} = 0, \label{Beq0}
\end{align} 
and the energy-momentum tensor can be expressed as  
\begin{align}
8 \pi G  \boldsymbol{T} = \frac{\tau_{+}(l)}{A(l)} (\boldsymbol{d}t  \otimes  \boldsymbol{d}t + \boldsymbol{d}l  \otimes  \boldsymbol{d}l) +  \frac{\tau_{-}(l)}{A(l)} (\boldsymbol{d}t  \otimes  \boldsymbol{d}l + \boldsymbol{d}l  \otimes  \boldsymbol{d}t),
\end{align}
with
\begin{align}
 \tau_{+}(l) 
 & := \tau_{v} \mathrm{e}^{2 l} + \tau_{u} \mathrm{e}^{- 2 l}  
 = A(l) + \frac{1}{2}\qty(A'(l)B'(l) - A''(l)), \label{tau+} \\
 \tau_{-}(l) 
&:
= \tau_{v}  \mathrm{e}^{2 l} - \tau_{u} \mathrm{e}^{- 2 l} 
=   A(l) B'(l) - \frac{1}{2}A'(l). \label{tau-}
\end{align}
By substituting Eqs.~\eqref{Aeq0} and \eqref{Beq0} into Eqs.~\eqref{tau+}  and \eqref{tau-}, we can see that $\tau_{u}$ and $\tau_{v}$ are constant.
Moreover, by using the freedom of the constant scaling of the coordinates $u$ and $v$,
which preserving the expression \eqref{expression of xi}, we can normalize the value of $\tau_{u}$ and $\tau_{v}$ arbitrarily.
In this paper,
we shall fix $\tau_u = \tau_v= - \frac{1}{4}$, assuming that both $\tau_{u}$ and $\tau_{v}$ are negative.
Then, we have
\begin{align}
 \tau_{+}(l) = - \frac{1}{2}\cosh(2 l), \quad
 \tau_{-}(l) = - \frac{1}{2}\sinh(2 l).
\end{align} 
We note the negative values of $\tau_u$ and $\tau_v$ imply a violation of the null energy condition.
We also note that fixing the scalings of the coordinates $u$ and $v$ removes the residual rescaling freedom, thereby rendering the normalizations (or origins) of $t$ and $l$ physically meaningful.

To summarize, with the self-similarity, the components of the Einstein equation reduce to
\begin{align}
 &B''(l)  +  \frac{\mathrm{e}^{B(l)}}{A(l)^{3/2}} = 0,\\
 &A''(l) - 4 A(l) - 2  \frac{\mathrm{e}^{B(l)}}{\sqrt{A(l)}} = 0, \\
 &2\tau_{+}(l) =  2A(l) + B'(l)A'(l) - A''(l) =- \cosh(2 l),\\
 &2\tau_{-}(l) = 2A(l) B'(l) - A'(l) =- \sinh(2 l).
\end{align}
In particular, the independent components can be written as follows:
\begin{align}
& A''(l)  - 2 A(l) - \frac{1}{2} \frac{A'(l)^2}{A(l)} -  \frac{A'(l) }{A(l)} \tau_{-}(l) + 2\tau_{+}(l) = 0, \label{Aeq} \\
&B(l) = \log \left(\frac{\sqrt{A(l)}}{2} \left(A''(l) - 4 A(l) \right) \right). \label{Beq}
\end{align}

Using these equations~\eqref{Aeq}, \eqref{Beq}, and the derivatives, we can compute the curvature invariants as
\begin{align}
 &R = 0, \\
 &R_{\mu\nu} R^{\mu\nu} = \frac{8\mathrm{e}^{-4t}}{L^4\qty[2A(\cosh(2l) - 2A
 ) + A'(l)(A'(l) - \sinh(2l))]^2}, \label{RR}\\
 &C_{\mu\nu\rho\sigma} C^{\mu\nu\rho\sigma} = \frac{12\mathrm{e}^{-4t}(2A\cosh(2l) -A'(l)\sinh(2l))^2}{L^4A^2 \qty[2A(\cosh(2l) - 2A)
 + A'(l)(A'(l) - \sinh(2l))]^2}. \label{CC}  
\end{align}

\subsection{Property of solutions}

In the previous subsection, under the homothetic symmetry, we have obtained the reduced ordinary differential equations.
In the following, we will analyze their solutions to determine whether they describe wormhole geometries.
For this purpose, we first establish throat conditions and then study asymptotics.

\subsubsection{Wormhole throat conditions}

In order to examine the geometry, we first discuss the structure of a wormhole.
For a spacetime to be qualified as a wormhole, it must possess a throat.
However, in non-static spacetimes, the definition of a wormhole throat is not unique, and several proposals can be found in the literature (see Refs.~\cite{Hochberg:1998ha,Hayward:1998pp,Maeda:2009tk,Tomikawa:2015swa} and also App.~\ref{throat}).
In the present self-similar setting, a natural prescription would be to define the throat as the surface of minimal areal radius on a constant-$t$ slice.
This choice is consistent with the homothetic symmetry, which selects constant-$t$ hypersurfaces as natural foliations.
Accordingly, we impose condition $A'(l_{\rm{th}})=0$ at the throat $l=l_{\rm{th}}$.
In addition, we require the flare-out condition $A''(l_{\rm{th}})>0$, which means that the surface $l=l_{\rm{th}}$ is not only extremal, but also minimal.
We note that the throat determined by this condition is compatible with the definition of Ref.~\cite{Maeda:2009tk} (See App.~\ref{throat}).

In this paper, we focus on the solutions which are symmetric under the inversion of the radial coordinate $l\rightarrow -l$.
In this case, a throat must be located at $l=0$.
Motivated by the above discussion, we shall impose the one parameter family of the initial conditions  for Eq.~\eqref{Aeq} which, we expect, corresponds to the symmetric wormhole solution,
\begin{align}
 A(0) = A_{\rm{th}} > 0, \qquad A'(0) = 0.
 \label{BC}
\end{align}
By evaluating Eq.~\eqref{Aeq} at $l=0$ with the conditions~\eqref{BC}, we obtain
\begin{align}
 A''(0) = 2 A_{\rm{th}} + 1 >0,
\end{align}
which means that the flare-out condition is automatically satisfied.
These conditions also mean that the function can be approximated as a quadratic function
\begin{align}
 A(l) \simeq A_{\rm{th}} + \qty(A_{\rm{th}} + \frac{1}{2})l^2 \label{Al2}
\end{align}
in the neighborhood of $l=0$.
From Eq.~\eqref{Beq}, we also find that
\begin{align}
 \frac{\mathrm{e}^{B(0)}}{\sqrt{A(0)}} = \frac{1}{2} - A_{\rm{th}}.
\end{align}
Thus, by requiring that $g_{tt} < 0$, that is, the vector $\boldsymbol{\partial}
_{t}$ is timelike, the value of the parameter $A_{\rm{th}}$ is restricted to
\begin{align}
 0 < A_{\rm{th}} < \frac{1}{2}. \label{rdomain}
\end{align} 

\subsubsection{Asymptotics and singularities}

In the previous subsection, we derived the form of the metric~\eqref{mettl} and the equations of motion~\eqref{Aeq} and \eqref{Beq}.
As discussed there, Eq.~\eqref{Aeq} is the only equation that needs to be solved.
However, it is difficult to obtain the general solution of the equation.
Therefore, in the next subsection we will solve Eq.~\eqref{Aeq} numerically.
Here, we examine the asymptotic behavior of the solution in the large-$|l|$ region.

One can show that 
\begin{align}
A_{0} := \frac{1}{4 \gamma} \mathrm{e}^{2 l} - \frac{1}{4 \gamma - 8} \mathrm{e}^{- 2 l} \label{Ainf}
\end{align}
is an exact solution of Eq.~\eqref{Aeq}.
Here, $\gamma$ is a constant.
Although $A_{0}(l)$ is an exact solution, we will mainly use it as an approximate solution in the large-$l$ region, where the first term dominates.
Therefore, we focus on $\gamma > 0$ to ensure $A(l) > 0$ asymptotically. 
As we will see below, the numerical solution suggests that $\gamma > 1$ 
as long as $A_{\rm{th}}$ satisfies~\eqref{rdomain}.
The metric derived from the solution~\eqref{Ainf} degenerates since
\begin{align}
\exp\qty(B_{0}(l)) := \sqrt{A_{0}} \left(A_{0}'' - 4 A_{0} \right) = 0,
\end{align}
where Eq.~\eqref{Beq} has been used.
Thus, $A_{0}$ is not viable as a physical solution.
Nevertheless, we may obtain physical geometries by considering perturbations around this degenerate solution.
Let us consider perturbations about~\eqref{Ainf} as follows:
\begin{align}
A(\epsilon;l) = A_{0}(l) + \epsilon \delta A(l) + \mathcal{O}(\epsilon^2),
\end{align}
assuming $\epsilon \delta A/A_{0} \ll 1$.
Hereafter, we focus on terms up to $O(\epsilon)$, and denote $\epsilon \delta A$ simply by $\delta A$.
Then, the linearized equation of motion for $\delta A$ can be expressed as
\begin{align}
\delta A'' + 
\left( \gamma - 2 + \frac{2 \gamma}{\gamma - (\gamma - 2) \mathrm{e}^{4l}} \right) \delta A'
+\left( - 2 \gamma + \frac{4 (\gamma - 1) \gamma }{\gamma - (\gamma - 2) \mathrm{e}^{4l}} \right) \delta A = 0.
\end{align}
Let us evaluate solution of this perturbation equation in the large-$l$ region.
In the regime where $\mathrm{e}^{4 l} \gg |\gamma/(\gamma - 2)|, \gamma/(\gamma - 2)^2$, we can approximate
the equation as
\begin{align}
\delta A'' + 
\left(\gamma - 2 - \frac{2 \gamma}{(\gamma - 2)} \mathrm{e}^{-4l}\right) \delta A'
+\left( - 2 \gamma - \frac{4 (\gamma - 1) \gamma }{ (\gamma - 2) }\mathrm{e}^{-4l} \right) \delta A = 0,
\end{align}
which has the solution
\begin{align}
\delta A &= \mathrm{e}^{-\gamma l} \mathrm{e}^{-\frac{\gamma}{2\gamma-4}\mathrm{e}^{-4l}}\qty(C_{1}U\qty(\frac{2+\gamma}{2}, \frac{6+\gamma}{4}, \frac{\gamma}{2(\gamma-2)}\mathrm{e}^{-4l
}) + C_{2}L_{-\frac{2+\gamma}{2}}^{\frac{2+\gamma}{4}}\qty(\frac{\gamma}{2(\gamma-2)}\mathrm{e}^{-4l
})
) .
\end{align}
Here, $U(a, b, c)$ and $L_{a}^{b}(c)$ denote the confluent hypergeometric function of the second kind and the generalized Laguerre function, respectively.
$C_{1}$ and $C_{2}$ are constants.
We see that, in the limit $l \rightarrow \infty$, the first term behaves as $\mathrm{e}^{2l}$, which is the same $l$-dependence as the first term in the background solution $A_{0}$.
Therefore, its contribution can be removed by a redefinition of $\gamma$, and we shall take $C_{1}$=0.
We also find that the second term can be approximated as
$L_{-\frac{2+\gamma}{2}}^{\frac{2+\gamma}{4}}\qty(\frac{\gamma}{2(\gamma-2)}\mathrm{e}^{-4l
}) = L_{-\frac{2+\gamma}{2}}^{\frac{2+\gamma}{4}}(0) + O\qty(\mathrm{e}^{-4l})$.
Then, for $\gamma\neq2$ and $l\rightarrow\infty$, we have the perturbed solution
\begin{align}
&A(l) = \frac{1}{4 \gamma} \mathrm{e}^{2 l} - \frac{1}{4 \gamma - 8} \mathrm{e}^{- 2 l} + \frac{C}{\gamma-2}\mathrm{e}^{-\gamma l}\qty(1 +O\qty(\mathrm{e}^{-4l})), \label{Ap} \\
&
\mathrm{e}^{B(l)} = \frac{\gamma + 2}{4 \sqrt{\gamma}} C \mathrm{e}^{(1-\gamma) l}\qty(1 +O\qty(\mathrm{e}^{-4l})),
 \label{Bp}
\end{align}
with $C$ being a positive constant.

\subsubsection{Global structure of the spacetime}
    
In the previous subsubsection, we derived the asymptotic form of the metric.
In order to discuss the structure of this spacetime, we evaluate null geodesics using the asymptotic solution.
Let us consider the future directed outgoing radial null curve of constant $t - l$ whose tangent vector is 
\begin{align}
\boldsymbol{k}=\frac{1}{-g_{tt}}\qty(\boldsymbol{\partial}_{t} + \boldsymbol{\partial}_{l}). \label{def k}
\end{align}
We see that this vector satisfies the affine parametrized geodesic equation $\nabla_{\boldsymbol{k}}\boldsymbol{k}=0$.
 The affine parameter $\lambda$ of this geodesic satisfies
\begin{align}
\frac{d l}{d \lambda} = k^{l} = \frac{1}{- g_{tt}}.
\end{align}
Therefore, in the limit $l \rightarrow\infty$ with $t-l$ being fixed, the affine parameter $\lambda$ of this geodesic is asymptotically written as
\begin{align}
\lambda = \int^{l} (-g_{tt})|_{t = l' + \text{const.}} d l' \propto - C \mathrm{e}^{ (2 - \gamma) l } + \text{const.}\notag
\end{align}
From this expression, for $\gamma<2$, the affine parameter diverges to $+\infty$ as $l \rightarrow\infty$.
Therefore, $l \rightarrow\infty$ with $t-l$ being fixed corresponds to the future null infinity for $\gamma<2$.
On the other hand, for $\gamma>2$, the affine parameter $\lambda$ remains finite even in the limit $l \to \infty$, which means that this limit is not actual infinity in terms of the proper distance.

Let us evaluate the curvature invariants in the asymptotic region.
As we discussed in the previous subsubsection, although the perturbative solution~\eqref{Ap} is not the exact solution of Eq.~\eqref{Aeq}, one may expect that this well describes the asymptotic behavior of $A(l)$ in terms of the behavior of the curvatures.
By substituting the asymptotic metric~\eqref{Ap} and \eqref{Bp} into the curvature invariants, we obtain
\begin{align}
 R_{\mu\nu} R^{\mu\nu} &\stackrel{l \to \infty}{\sim}  \mathrm{e}^{- 4t} \mathrm{e}^{2(\gamma-2)l} =   \mathrm{e}^{- 4(t - l)} \mathrm{e}^{2(\gamma - 4 ) l}
 =  \mathrm{e}^{- 4(t + l)} \mathrm{e}^{2\gamma l}
 , \label{RRas} \\
&  C_{\mu\nu\rho\sigma} C^{\mu\nu\rho\sigma} \stackrel{l \to \infty}{\sim}  \gamma^2 \mathrm{e}^{- 4t} \mathrm{e}^{2(\gamma-4)l} . \label{CCas}
\end{align}
Here, the symbol “$\sim$” denotes asymptotic equivalence up to a numerical prefactor.
From these equations, we see that
the spatial infinity ($l \rightarrow\infty$ with $t$ being fixed) has finite curvature invariants if $\gamma < 2$ and has divergent ones if $\gamma > 2$.
Similarly, the future null infinity ($l \to \infty$ with $t - \ell$ being fixed) has finite curvature invariants if $\gamma < 4$ and divergent ones for $\gamma > 4$.
Thus, in the region of $2<\gamma<4$, the curvature scalars converges in the future null direction, while the null geodesic is incomplete (Table~\ref{tbgamma}).

\begin{table}
 \begin{center}
\begin{tabular}{|c|c|c|c|c|} \hline
& null geodesic & \makecell{curvature scalar\\at spatial infinity} & \makecell{curvature scalar\\at future null infinity} & \makecell{curvature scalar\\at past null infinity} \\ \hline
 $\gamma<2$   & complete   & finite & finite & diverge\\ \hline
 $2<\gamma<4$ & incomplete & diverge
 & finite & diverge\\ \hline
 $4<\gamma$   & incomplete & diverge & diverge & diverge \\ \hline
\end{tabular}
  \caption{The dependence of the behavior of the null geodesic and the curvature scalar on the value of $\gamma$.}
    \label{tbgamma}
 \end{center}
 \end{table}

To investigate the regularity for $2<\gamma<4$, we shall discuss the extendibility of the geometry in the null direction.
In order to do that, we shall compute the components of the curvature tensor expanded in parallelly propagated null tetrad.
From the above computation, we see that
\begin{align}
&\boldsymbol{e}^{+} = \frac{1}{\sqrt{2}}\qty(\boldsymbol{d}t - \boldsymbol{d}l), \\
&\boldsymbol{e}^{-} = \mathrm{e}^{2 t}L^2
\frac{\mathrm{e}^{B(l)}}{\sqrt{2A(l)}}\qty(\boldsymbol{d}t +
\boldsymbol{d}l), \\
&\boldsymbol{e}^{\theta} = \mathrm{e}^{t}L\sqrt{A
(l)}\boldsymbol{d}\theta, \\
&\boldsymbol{e}^{\phi} = \mathrm{e}^{t}L\sqrt{A(l)}\sin\theta \boldsymbol{d}\phi,
\end{align}
are parallelly propagated along the radial null geodesics in the large-$l$ region.
Namely, they satisfy

\begin{align}
\bm{\nabla}_{\bm{k}} \bm{e}^{+} = 0,~
\bm{\nabla}_{\bm{k}} \bm{e}^{-} = 0,~
\bm{\nabla}_{\bm{k}} \bm{e}^{\theta} = 0,~
\bm{\nabla}_{\bm{k}} \bm{e}^{\phi} = 0,
\end{align}
and
\begin{align}
\boldsymbol{g} = -\qty(\boldsymbol{e}^{+}\otimes \boldsymbol{e}^{-} + \boldsymbol{e}^{-}\otimes \boldsymbol{e}^{+}) + \boldsymbol{e}^{\theta} \otimes \boldsymbol{e}^{\theta} + \boldsymbol{e}^{\phi} \otimes \boldsymbol{e}^{\phi}
\end{align}
The non-zero components in this basis behave as
\begin{align}
 &R^{\mu\nu} e_{\mu}^{+}e_{\nu}^{+} \stackrel{l \to \infty}{\sim} \mathrm{e}^{2(\gamma-2)l - 4(t-l)}, \label{Rppas} \\
&R^{\mu\nu} e_{\mu}^{-}e_{\nu}^{-}  \stackrel{l \to \infty}{\sim} \mathrm{e}^{-4l}, \label{Rmmas} \\
&C^{\mu\nu\rho\sigma} e_{\mu}^{+}e_{\nu}^{-}e_{\rho}^{+}e_{\sigma}^{-} \stackrel{l \to \infty}{\sim} 
\mathrm{e}^{-2(t-l)}\mathrm{e}^{(\gamma-6)l}, \\
&C^{\mu\nu\rho\sigma}  e_{\mu}^{\theta}e_{\nu}^{\phi}e_{\rho}^{\theta}e_{\sigma}^{\phi} \stackrel{l \to \infty}{\sim}   
\mathrm{e}^{-2(t-l)}\mathrm{e}^{(\gamma-6)l}, \\
&C^{\mu\nu\rho\sigma} e_{\mu}^{+}e_{\nu}^{\theta}e_{\rho}^{-}e_{\sigma}^{\theta}
= C^{\mu\nu\rho\sigma} e_{\mu}^{+}e_{\nu}^{\phi}e_{\rho}^{-}e_{\sigma}^{\phi} \stackrel{l \to \infty}{\sim}  
\mathrm{e}^{-2(t-l)}\mathrm{e}^{(\gamma-6)l}.
\end{align}
From these relations, we find that the component of the curvature tensor $R^{\mu\nu} e_{\mu}^{+}e_{\nu}^{+}$ diverges as $l \rightarrow\infty$ with $t-l$ being fixed for $2<\gamma$.
Therefore, there is parallelly propagated (pp)-curvature singularity~\cite{Hawking:1973uf,Clarke:1973,Ellis:1977pj,Clarke:1982} in the future null direction when $2<\gamma$.
In terms of the energy-momentum tensor, this singular component is written as
\begin{align}
 &T^{\mu\nu} e_{\mu}^{+}e_{\nu}^{+} \stackrel{l \to \infty}{\sim} \mathrm{e}^{2(\gamma-2)l - 4(t-l)}\tau_{v} \label{Tppas}.
\end{align}
This implies that the existence of the pp-curvature singularity for the Ricci tensor in the future null region is due to the infinite flux of the negative energy in the $v$-direction i.e. out-going future null direction.
Similarly, we can show that the flux of the negative energy in the $u$-direction diverges in the limit $l\rightarrow -\infty$ with $t + l$ being fixed.

Next, we shall evaluate the behavior of the null geodesic and the curvature scalar in the past direction.
From Eqs.~\eqref{RRas} and \eqref{CCas}, it is obvious that there is scalar curvature singularity in the past timelike direction ($t\rightarrow - \infty$ with $l$ being fixed).
Similar to the future case, we see that the past-directed outgoing vector $\boldsymbol{\tilde{k}}=\qty(-\boldsymbol{\partial}_{t} + \boldsymbol{\partial}_{l})/(-g_{tt})$ satisfies the geodesic equation $\nabla_{\boldsymbol{\tilde{k}}}\boldsymbol{\tilde{k}}=0$.
Therefore, the affine parameter $\tilde{\lambda}$ can be written as
\begin{align}
 \tilde{\lambda} &\stackrel{l \to \infty}{\sim} - \mathrm{e}^{-(2 + \gamma)l} + \mathrm{const.},
\end{align}
which approach to a constant exponentially with $l$ as long as $\gamma>0$.
From Eqs.~\eqref{RRas} and \eqref{CCas}, curvature invariants behaves as $R_{\mu\nu} R^{\mu\nu} \sim\mathrm{e}^{2\gamma l}$ and $C_{\mu\nu\rho\sigma} C^{\mu\nu\rho\sigma}  \sim\mathrm{e}^{2\gamma l}$ in the limit $l \rightarrow \infty$ with $t+l$ being fixed, and diverge in the asymptotic region.
Therefore, in the past null direction, there is scalar curvature singularity regardless of the value of $\gamma$.

\begin{figure}[H]
  \centering
\includegraphics[width=0.8
\textwidth]{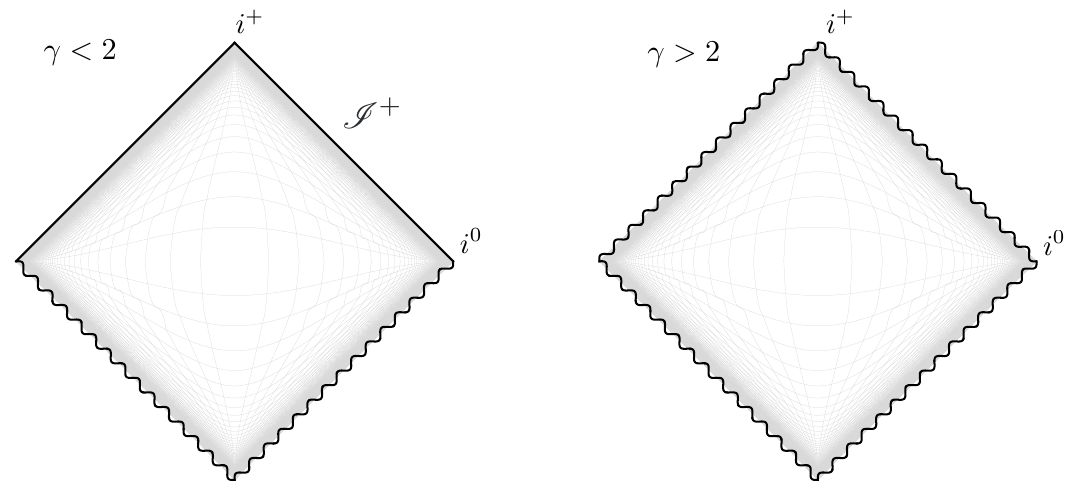}
  \caption{The Penrose diagrams of the geometry for $\gamma<2$ (left panel) and $\gamma>2$ (right panel).
  For $\gamma<2$, the geometry is free from singularity in the spatial and the future null infinity, while there is singularity for $\gamma>2$.
  In the past region, there is curvature singularity regardless of the value of $\gamma$.
  The figure also includes $t$-constant and $l$-constant lines, drawn as gray curves.
  }
  \label{fig:Penrose}
\end{figure}

Combining the analysis at the throat and asymptotics, the global picture can be summarized as follows:
\begin{itemize}
    \item At the throat, the geometry is regular for $0<
    A_{\rm{th}}<1/2$.
    \item For $\gamma<2$, both spatial and future null infinities are complete and regular.
    \item For $\gamma>2$, future null geodesics terminate at finite affine parameter (pp-curvature singularity), and for $\gamma>4$, scalar invariants also diverge.
    \item In the past direction, all solutions possess an unavoidable singularity at $t\rightarrow -\infty$.
\end{itemize}
The global structure of the geometry is expressed as Fig.~\ref{fig:Penrose}.

\subsection{Numerical solution}

In this subsection, to confirm the analytic expectations derived above,
we integrate the ordinary differential equations numerically.

\begin{figure}[H]
  \centering  \includegraphics[width=0.8\textwidth]{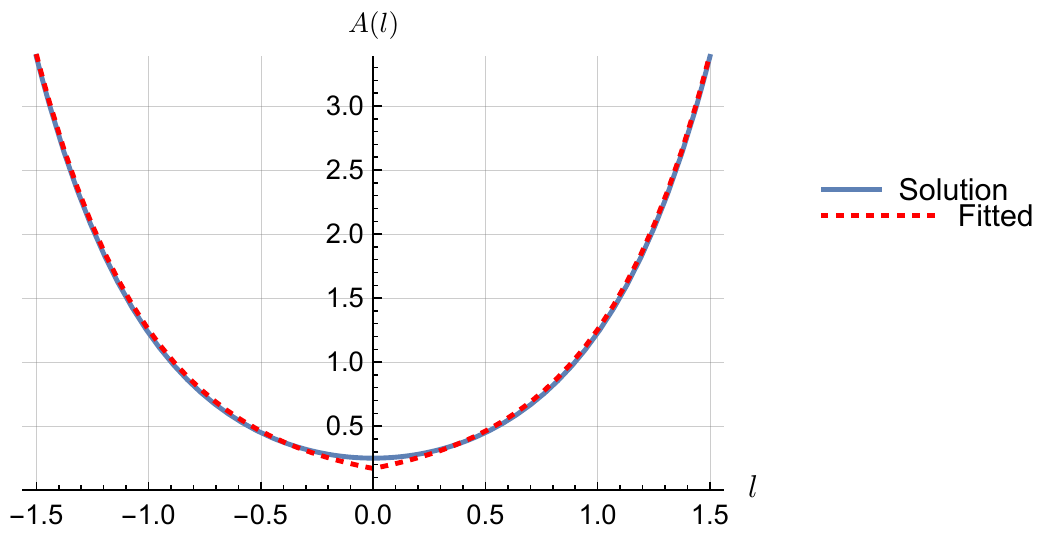}
  \caption{Plot of the solution of Eq.~\eqref{Aeq} with $A_{\rm{th}}=1/4$. (blue solid curve).
  The red dashed curve is the fitted line $\mathrm{e}^{2|l|}/4 \gamma$ with $\gamma \simeq 1.474$.
  $A(l)$ takes minimum at $l=0$, and increases monotonically with as $l$ increases.}
  \label{fig:Br}
\end{figure}

The numerical solution of Eq.~\eqref{Aeq} with the boundary condition~\eqref{BC} is depicted in Fig.~\ref{fig:Br}.
We have fixed the throat radius $A_{\rm{th}}=1/4$.
We see that, the area at fixed time, $A(l)$, increases monotonically with the absolute value of the radial coordinate $l$, confirming
the presence of a wormhole throat as predicted by the analytic throat 
conditions~\eqref{BC}.

\begin{figure}[H]
  \centering
  \includegraphics[width=0.6\textwidth]{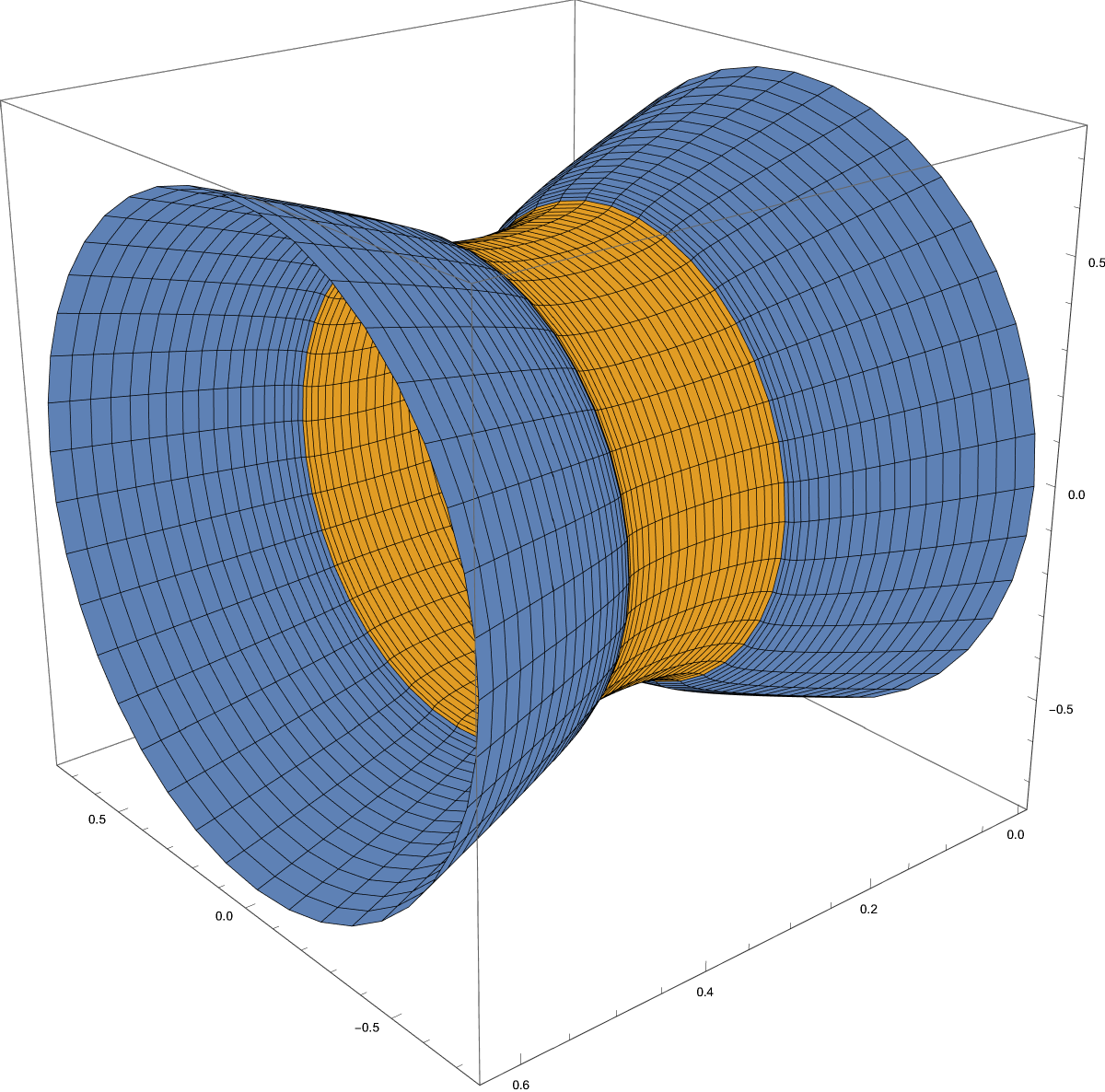}
  \caption{The embedded diagram of the geometry around the throat at some moment ($A_{\rm{th}}=1/4$) into the three-dimensional spaces.
  The orange region is embedded geometry in the Euclid space, while the blue colored region is the Minkowski spacetime.}
  \label{fig:EBD}
\end{figure}

To provide a geometric visualization, Fig.~\ref{fig:EBD} displays the corresponding embeddings of constant-time slices into three-dimensional flat space.
The orange region is embedded in the three-dimensional Euclidean space, while the gray region is embedded in the Minkowski spacetime.
This visualization clearly illustrates the throat structure in geometric terms.
These embedding diagrams make the wormhole structure manifest: the throat at $l=0$ appears as the minimal surface connecting two asymptotic regions.

\begin{figure}[H]
  \centering
\includegraphics[width=0.6\textwidth]{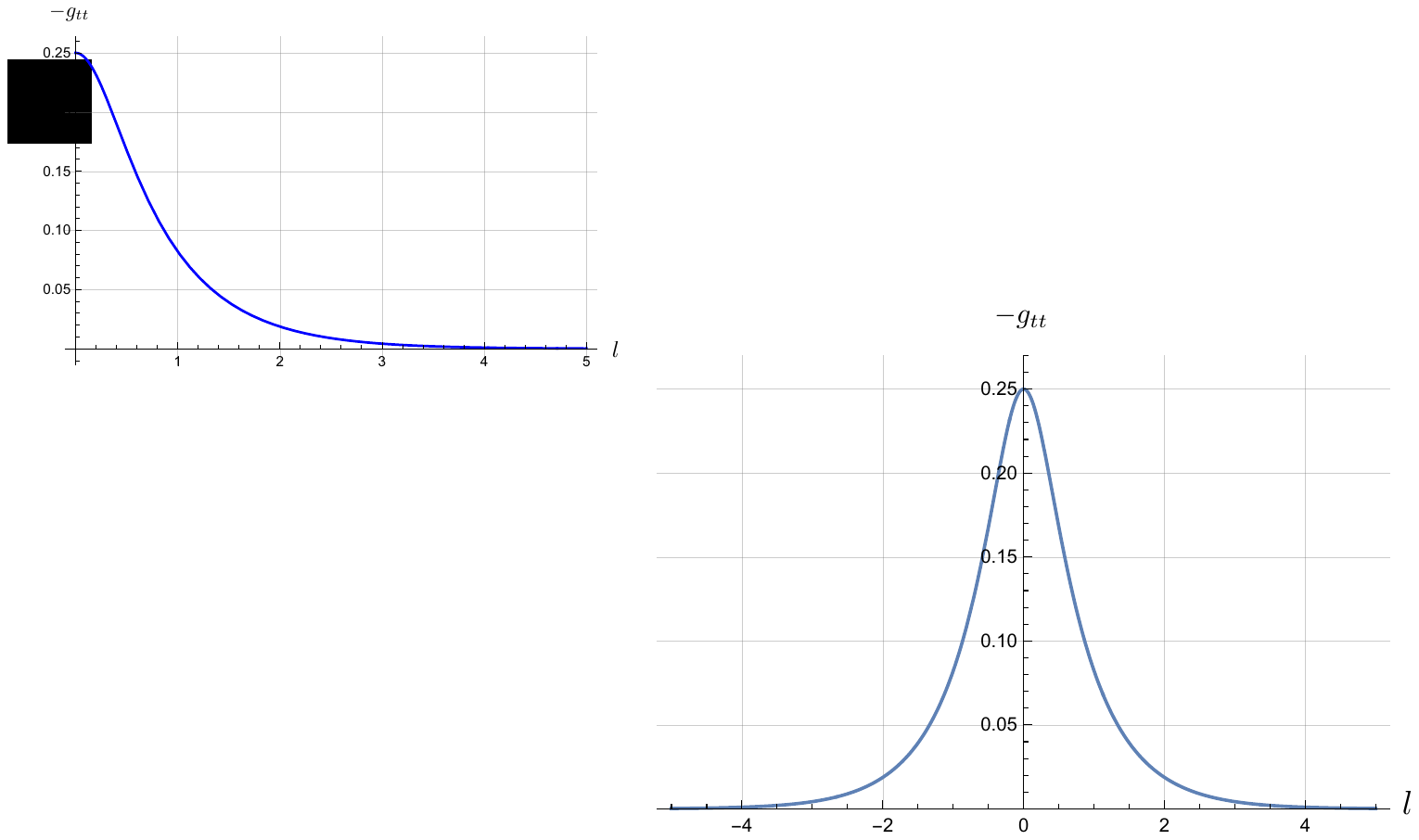}
  \caption{Plot of $- g_{tt}/\mathrm{e}^{2t}L^2 = \mathrm{e}^{B(l)}/\sqrt{A(l)}$ for the wormhole solutions with $A_{\rm{th}} = 1/4$. The solution satisfy $- g_{tt} > 0$ throughout the entire domain. }
  \label{fig:gtt}
\end{figure}

The $l$-dependence of the metric component $g_{tt}/\mathrm{e}^{2t}L^2=\mathrm{e}^{A(l)}/\sqrt{B(l)}$ is shown in Fig.~\ref{fig:gtt}.
The figure shows that $g_{tt}<0$ holds everywhere, confirming that $t$ is a timelike coordinate throughout the entire domain.

\begin{figure}[H]
  \centering
\includegraphics[width=0.6\textwidth]{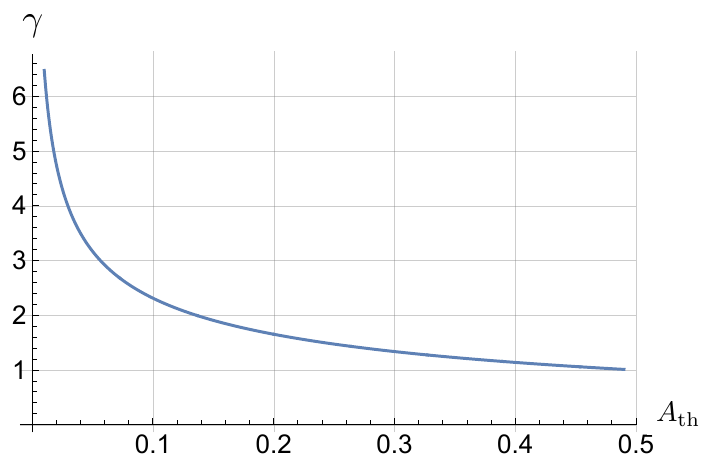}
  \caption{The value of fitted parameter $\gamma$ in Eq.~\eqref{Ainf}.
  $\gamma$ decreases as the throat radius $A_{\rm{th}}$ increases, and asymptotes to $\gamma=1$ as $A_{\rm{th}}$ approaches to $1/2$.}
  \label{fig:gamma}
\end{figure}

In Fig.~\ref{fig:Br}, we also plot the fitted curve corresponding to the growing part of the singular solution~\eqref{Ainf}, $A_{0} \simeq  \mathrm{e}^{2 |l|}/4\gamma$, by the red dotted line.
For $A_{\rm th}=1/4$, the fitted parameter is $\gamma\simeq 1.474$.
This figure indicates that the behavior of $A(l)$ in the large-$l$ region is well approximated by the exponential form.
This agreement demonstrates the consistency between the analytic asymptotics and the numerical results in the large-$l$ region.
We note, however, there is discrepancy between the asymptotic approximation and the numerical solution near the throat.
The dependence of the value of the fitted parameter $\gamma$ on the throat radius $A_{\rm{th}}$ is shown in Fig.~\ref{fig:gamma}.
As $A_{\rm{th}}$ increases toward the upper bound $1/2$, the fitted value of $\gamma$ decreases monotonically and asymptotes to  $\gamma=1$.
As discussed in the previous subsection, the geometry remains regular at spatial infinity and future null infinity if and only if $\gamma<2$.
Accordingly, we find that no singularity appears in the spatial or future null directions for wormholes with sufficiently large throats ($A_{\rm th}>0.14$).

\section{Formation of wormhole}
\label{Formation}

In Sec.~\ref{SSWS}, we constructed the  self-similar wormhole solution.
We now turn to the question of how such geometries can form dynamically.
To this end, we construct a composite spacetime by  cutting and gluing together exact solutions of Einstein’s equations across null shells, generalizing the scenario proposed by Hayward and Koyama~\cite{Hayward:2004wm,Koyama:2004uh}.

\subsection{Overall strategy}
\label{Strategy}

Our aim is to construct a dynamical scenario in which an initial 
Schwarzschild black hole is converted into a self-similar wormhole 
by irradiation of negative-energy flux. To achieve this, we patch 
together three distinct exact solutions of Einstein’s equations:
\begin{enumerate}
    \item  Schwarzschild region: the initial state of the black hole prior to the injection of the negative energy. 
    This serves as the starting point of the formation process.

 \item Negative-energy Vaidya regions: the intermediate states in which the ingoing and outgoing streams of null radiation carrying negative energy. 
 These regions mediate the transition between the Schwarzschild black hole and the 
   wormhole geometry.

 \item Self-similar wormhole region: the final state described by the solutions constructed in the previous section. This region contains the wormhole throat and its asymptotic domains.
\end{enumerate}

\begin{figure}[H]
  \centering
  \includegraphics[scale = 0.7]{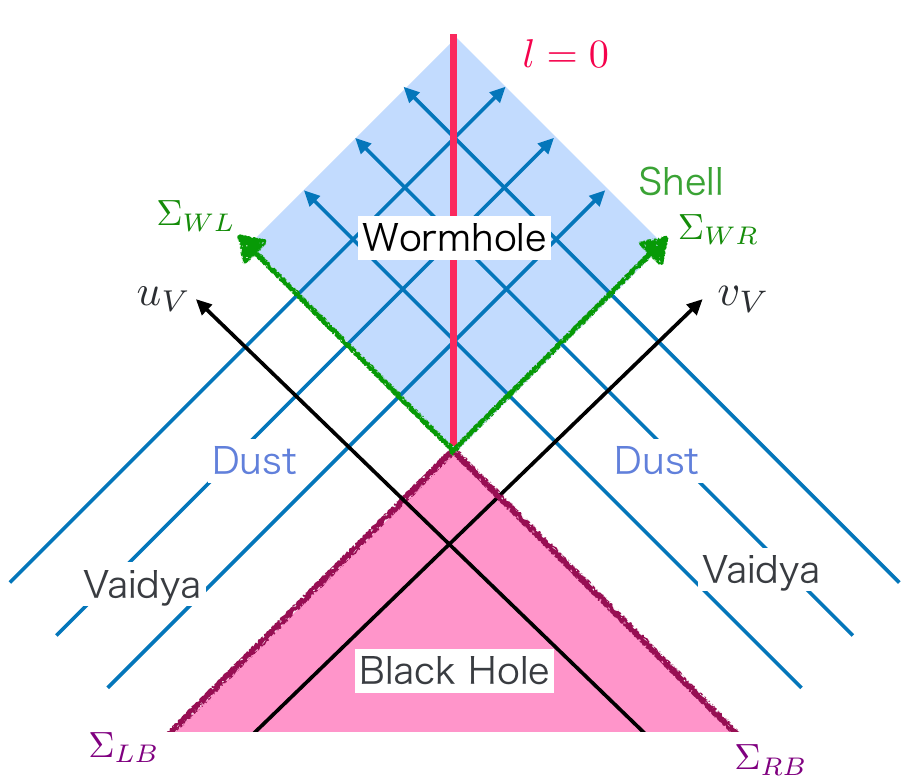}
  \caption{The schematic picture of the wormhole formation model. 
  We start from a black hole, and go to a wormhole via Vaidya regions. 
  The regions are separated by the null shells without pressure, which are shown as purple lines and green arrows.}
  \label{Schem}
\end{figure}

The overall structure is illustrated schematically in Fig.~\ref{Schem}. 
The spacetime is divided into patches separated by null hypersurfaces, 
across which the geometries are matched using the Barrabès–Israel 
junction formalism~\cite{Barrabes:1991ng}. 
For simplicity, we assume the junctions are left–right symmetric.
At each junction, thin null shells appear, whose 
energy–momentum tensors encode the effect of the negative-energy flux of the shells.
In this model, considering continuity of injected energy, we assume that the shell has no pressure but only energy.

In the next subsection, we describe each building block in turn: 
the Schwarzschild region, the Vaidya regions, 
and the wormhole region.
Then, we apply the 
junction conditions to derive the relations among the throat radius, 
the injected flux, and the initial black-hole parameters, ensuring 
consistency of the whole construction.

\subsection{Geometries}
\label{Geo}

\subsubsection{Schwarzschild black hole}

The vacuum solution of Einstein equation with static and spherically symmetry is known as Schwarzschild black hole:
\begin{align}
  \boldsymbol{g} &= - f_B(r) \boldsymbol{d}t \otimes \boldsymbol{d}t  + \frac{1}{f_B(r)} \boldsymbol{d}r \otimes \boldsymbol{d}r  + r^2 \boldsymbol{d}\Omega^2, \\
  f_B(r) &= 1 - \frac{2M}{r}. 
\end{align}
Here, $M$ represents the mass, and there is a horizon at $r=2M $.  

For later calculation, we shall express the metric in the Eddington-Finkelstein coordinates.
If we consider the region which will be patched with the right Vaidya region, the spacetime can be written
as
\begin{equation}
  \begin{aligned}
      \boldsymbol{g} &= -  f_B \boldsymbol{d}v_{B} \otimes \boldsymbol{d}v_{B}  + \boldsymbol{d}v_{B}\otimes  \boldsymbol{d}r  + \boldsymbol{d}r \otimes  \boldsymbol{d}v_{B} + r^2 \boldsymbol{d}\Omega^2. 
  \end{aligned}
\label{efc}
\end{equation}
Here, $v_{B}$ is the ingoing null coordinate defined by $v_{B} := t + r_*$, with $r_* := \int \dd{r} 1 / f_{B}(r)$ being a tortoise coordinate. 

\subsubsection{Vaidya spacetime}

Next, to describe the intermediate state from a black hole through the injection of negative energy to a wormhole, we introduce two-sided regions described by Vaidya spacetimes.
The Vaidya spacetimes may describe spherical accretion process of the energy onto stars and black holes.

For the right Vaidya region, which is patched with the wormhole geometry with $l>0$ the region, we write the metric as
\begin{equation}
  \boldsymbol{g} = - f_{R}(v_{V}, r) \boldsymbol{d}v_{V}\otimes\boldsymbol{d}v_{V}  + \boldsymbol{d}v_{V}\otimes \boldsymbol{d}r + \boldsymbol{d}r \otimes \boldsymbol{d}v_{V} + r^2 \boldsymbol{d}\Omega^2 . \label{vaidyaR}
\end{equation}
Here, $v_{V}$ denotes the ingoing null coordinate and
\begin{equation}
  f_{R}(v_{V}, r) := 1- \frac{2m(v_{V})}{r},
\end{equation}
where $m(v_{V})$ is a mass function.
The energy-momentum tensor is given by
\begin{align}
  \boldsymbol{T} = \frac{1}{4 \pi r^2} \dv{m(v_{V})}{v_{V}}  \boldsymbol{d}v_{V}\otimes\boldsymbol{d}v_{V}. 
\end{align}
We assumed that only the mass $m$ would be changed by the negative-energy null dust $\qty(m'(v_{V}) < 0)$. 

For later convenience, we rewrite the line element in the same form as in Eq.~\eqref{efc};
\begin{equation}
  \begin{aligned}
 \boldsymbol{g} &= -  f_R \boldsymbol{d}v_{V}\otimes\boldsymbol{d}v_{V} + \boldsymbol{d}v_{V}\otimes  \boldsymbol{d}r  + \boldsymbol{d}r\otimes  \boldsymbol{d}v_{V} + r^2 \boldsymbol{d}\Omega^2.
  \end{aligned}
\label{vaiRefc}
\end{equation}

For the left Vaidya region, glued with the wormhole region in $l<0$, we shall also introduce the metric using the Eddington-Finkelstein coordinates as
\begin{equation}
  \begin{aligned}
  \boldsymbol{g} &= - f_L \boldsymbol{d}u_{V}\otimes\boldsymbol{d}u_{V} + \boldsymbol{d}u_{V}\otimes  \boldsymbol{d}r  + \boldsymbol{d}r\otimes  \boldsymbol{d}u_{V} + r^2 \boldsymbol{d}\Omega^2, \\
    f_L &= 1 - \frac{2m(u_{V})}{r}.
  \end{aligned}
\label{vaiLefc}
\end{equation}
We note that, in the left region, $u_{V}$ is used as the ingoing null coordinate, instead of $v_{V}$.
The energy-momentum tensor is given by
\begin{align}
  \boldsymbol{T} = \frac{1}{4 \pi r^2} \dv{m(u_{V})}{u_{V}} \boldsymbol{d}u_{V}\otimes\boldsymbol{d}u_{V}
\end{align}
with $m'(u_{V}) < 0$.

\subsubsection{Traversable wormhole}

The geometry of the traversable wormhole is what we discussed in the previous section. 
We write the metric in double null coordinates as
\begin{align}
 \boldsymbol{g} = - \frac{1}{2} \frac{\mathrm{e}^{B \left(\frac{v}{u}\right)}}{\sqrt{A\left(\frac{v}{u}\right)}} \left(\boldsymbol{d}u \otimes \boldsymbol{d}v +\boldsymbol{d}v \otimes \boldsymbol{d}u \right) + u v A\left(\frac{v}{u}\right) \boldsymbol{d}\Omega^2.
 \label{WHuv}
\end{align}

\subsection{Junction conditions}
\label{Junction}

We will describe the construction by matching the three building blocks introduced above across null hypersurfaces. 
In this subsection, we review the Barrabès–Israel formalism~\cite{Barrabes:1991ng}, which provides the junction conditions for null shells in general relativity.

In order to express the junction conditions in terms of intrinsic quantities of the null hypersurfaces, we introduce some notations.
Consider patching two geometries across thin null shell
$\Sigma$.
We shall express the position of the shell using the equation $\Phi(x^{\mu})=0$.
Let $\xi^{a}$ $(a=1,2,3)$ be intrinsic coordinates on the shell, and define the tetrad $e^{\mu}_{(a)} := \partial x^{\mu}/\partial \xi^{a}$, where $x^{\mu}$ are the coordinates of the four-dimensional spacetime.
Since the choice of $\xi^{a}$ is not unique, there is a corresponding ambiguity in $e^{\mu}_{(a)}$.
Using this ambiguity, we shall fix the tetrad such that $\boldsymbol{e}_{(1)}=\boldsymbol{n}^{\sharp}$, where $\boldsymbol{n}$ is a null normal to the shell. 
Here, the symbol $\sharp$ denotes the metric dual (index-uppering) map by the metric $\bm{g}$, that is, $\bm{n}^{\sharp} = g^{\mu\nu}n_{\mu} \bm{\partial}_{\nu} = n^{\mu}  \bm{\partial}_{\mu}$.
Analogously, $\flat$ will denote the index-lowering operation, defined by $\bm{V}^{\flat} = g_{\mu\nu}V^{\mu} \bm{d} x^{\nu} = V_{\mu} \bm{d}x^{\mu}$ for a given vector field $\bm{V}$.
We shall denote the rest of the tetrad as $\boldsymbol{e}_{(A)}$ $(A=2,3)$.
Then we introduce another null vector $\boldsymbol{N}$, which is called the transverse vector, satisfying $\boldsymbol{n}^{\sharp} \cdot \boldsymbol{N} = -1$ and $\boldsymbol{e}_{(A)} \cdot 
\boldsymbol{N}=0$ $(A=2,3)$.
Here, the dot denotes the inner product of two vectors with respect to the metric $\boldsymbol{g}$.
Using this definition of $\boldsymbol{N}$, the normal $\boldsymbol{n}$ can be expressed as
  \begin{align}
    \boldsymbol{n} &=-\frac{1}{\alpha}\boldsymbol{d}\Phi, \label{ndPhi} \\
    \alpha &= \nabla_{\boldsymbol{N}}\Phi.
\end{align}

Using these tetrads and vectors, the induced metric and the transverse extrinsic curvature of $\Sigma$ can be defined as
  \begin{align}
&h_{AB}:=g_{\mu\nu}e^{\mu}_{(A)}e^{\mu}_{(B)}, \\
&\mathcal{R}_{ab}:=-N_{\mu}e^{\nu}_{(b)}\nabla_{\nu}e^{\mu}_{(a)}.
\end{align}
As a practical rule throughout this section, we will choose intrinsic coordinates $\xi^{a}$ so that $(\xi^{2}, \xi^{3}) =(\theta, \phi)$ parametrize the spherical sections. 
This choice keeps $h_{AB}$ manifestly continuous and diagonal and simplifies $\mathcal{R}_{ab}$ blocks.

The spacetime under consideration is spherically symmetric, and we perform the patching along spherically symmetric null hypersurfaces.
The junction conditions for spherical null hypersurfaces are written as follows;
The first junction condition imposes the continuity of the induced metric 
  \begin{equation}
    [h_{AB}]  = 0 . \label{firjc}
  \end{equation}
  Here, we have introduced the expression for the difference of quantity $X$ across the hypersurface by $[X] := X|_{\Phi \rightarrow +0} - X|_{\Phi \rightarrow -0}$. 
In spherical symmetry, the angular part of the induced metric is proportional to $r^2$, where $r$ is the areal radius.
  Therefore, condition \eqref{firjc} ensures that the areal radius is continuous across the hypersurface, allowing us to introduce a consistent radial coordinate $r$ on it.

The second junction condition identifies the difference in the extrinsic curvature and the energy-momentum of the shell.
Concretely, the surface stress tensor $\boldsymbol{T}_{\Sigma}$ is identified with the discontinuity of the transverse curvature
  \begin{align}
        \boldsymbol{T}_{\Sigma} &= \alpha \boldsymbol{S} \delta (\Phi), \\
    \boldsymbol{S} &= \sigma \boldsymbol{n} \otimes \boldsymbol{n}  + j^A \qty(\boldsymbol{n} \otimes \boldsymbol{e}^{\flat}_{(A)} + \boldsymbol{e}^{\flat}_{(A)} \otimes \boldsymbol{n}) + P h^{AB} \boldsymbol{
e}^{\flat}_{(A)} \otimes \boldsymbol{e}^{\flat}_{(B)}, \\
    \sigma &= - \frac{1}{8 \pi} h^{AB} \qty[\mathcal{R}_{AB}], \quad
    j^A = \frac{1}{8 \pi} h^{AB} \qty[\mathcal{R}_{1 B}], \quad
    P = - \frac{1}{8 \pi} \qty[\mathcal{R}_{1 1}]. \label{2nd}
\end{align}
Here, $\sigma$, $j^{A}$, and $P$ denote the energy density, current and the pressure of the shell, respectively.

\subsection{Energy-momentum tensor of the shell}
\label{EMS}

In this subsection, we will compute the energy-momentum tensors of the shells, imposing the junction conditions and the pressureless conditions.

Before applying the junction conditions in detail, let us specify the
null shells that separate the different regions of the spacetime. 
We denote these shells by $\Sigma_{XY}$, where $X$ and $Y$ label the
adjacent regions being joined. 
Here, the black hole, right Vaidya, left Vaidya, and wormhole regions are labeled as $B$, $R$, $L$, and $W$, respectively.
In what follows, we focus on the right
hand side of the spacetime diagram (See Fig.~\ref{Schem}).

In the right region, the location of the first shell, $\Sigma_{RB}$ is specified by the scalar equation $v_{B} = v_{B,0}$ (constant) in the black hole region, and by $v_{V} = v_{V,0}$ (constant) in the right Vaidya region.
In contrast, the second shell $\Sigma_{WR}$ is located at $v_{V} = \zeta_{V}(r)$ in the right Vaidya region, and lies along the surface $u = u_{0}$ (constant) in the wormhole geometry.
The function $\zeta_{V}(r)$ can be derived from the Vaidya metric~\eqref{vaiRefc} by solving the equation
\begin{align}
 &\dv{\zeta_{V}(r)}{r} = \frac{2}{f_{R}(\zeta_{V}(r), r)}.
 \label{zetaR}
\end{align}

\subsubsection{Matching black hole and Vaidya}

\begin{figure}[H]
  \centering
  \includegraphics[scale = 0.5]{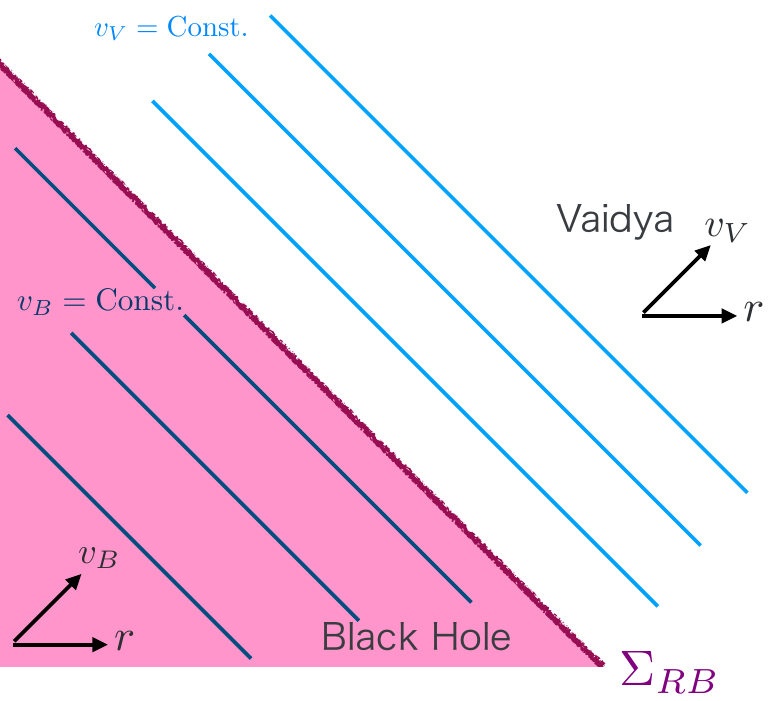}
  \caption{The picture of the black hole region which is described by Eq.~\eqref{efc}, and the right Vaidya region which is described by~\eqref{vaiRefc} in the vicinity of $\Sigma_{RB}$.
  The position of the shell $\Sigma_{RB}$ is described by the equations $v_{B}-v_{B,0}=0$ and $v_{V}-v_{V,0}=0$ from the black hole region and the Vaidya region, respectively.}
  \label{SchemRB}
\end{figure}

Firstly, let us discuss the junction of the black hole region and the right Vaidya region across the shell $\Sigma_{RB}$ (Fig.~\ref{SchemRB}).
As discussed above, in the black hole region, $\Sigma_{RB}$ is determined by the equation $\Phi^{RB-B}=v_{B}-v_{B,0}=0$.
Here, the superscript $RB-B$ indicates that the quantity is evaluated in the black hole region ($B$) for the junction across $\Sigma_{RB}$.
In what follows, $XY-Y$ will refers to a quantity evaluated in region $Y$ for the junction across $\Sigma_{XY}$.
By choosing the intrinsic
coordinates as $\xi^{a}=(-r, \theta, \phi)$, we have
\begin{equation}
  \boldsymbol{e}^{RB}_{(1)} 
 = -\boldsymbol{\partial}_{r} 
, \quad \boldsymbol{e}^{RB}_{(2)} = \boldsymbol{\partial}_{\theta}, \quad \boldsymbol{e}^{RB}_{(3)} =\boldsymbol{\partial}_{\varphi}. 
\end{equation}
The normal $\boldsymbol{n}^{RB}$ is written as
\begin{align}
   \boldsymbol{n}^{RB} 
   = -\boldsymbol{d}v_{B}, \label{normaB} 
\end{align}
and the transverse null vector is
\begin{align}
  \boldsymbol{N}^{RB-B , \flat}&= - \frac{f_{B}}{2} \boldsymbol{d}v_{B} + \boldsymbol{d}r. 
\end{align}
Then, the transverse extrinsic curvature is
\begin{align}
  \mathcal{R}^{RB-B}_{ab}
  & = \mathrm{diag} \qty(0,  \frac{rf_{B}}{2}, \frac{rf_{B}}{2} \sin^2 \theta),
  \label{exuB}
\end{align}
and the parts which contribute to the components of the surface stress tensor are
  \begin{align}
    h^{AB} \mathcal{R}^{RB-B}_{AB} &= \frac{f_{B}}{r}, \quad
    {R}^{RB-B}_{1 B}=0 , \quad
    \mathcal{R}^{RB-B}_{1 1} = 0.  \label{exuB2}
\end{align}

From the Vaidya region, $\Sigma_{RB}$ is determined by the equation $\Phi^{RB-R}=v_{V}-v_{V,0}=0$.
Similar to the black hole case, we have
\begin{align}
  \boldsymbol{n}^{RB} &= -\boldsymbol{d}v_{V}, \label{normaR}\\
  \boldsymbol{N}^{RB-R, \flat} &= - \frac{f_{R}}{2} \boldsymbol{d}v_{V} + \boldsymbol{d}r.
  \end{align}
Therefore, we obtain
  \begin{align}
  \mathcal{R}^{RB-R}_{ab} &= \mathrm{diag} \qty(0,  \frac{rf_{R}}{2}, \frac{rf_{R}}{2} \sin^2 \theta),
  \label{exuR}
\end{align}
and
  \begin{align}
    h^{AB} \mathcal{R}^{RB-R}_{AB} &= \frac{f_{R}}{r}, \quad
    {R}^{RB-R}_{1 B}=0 , \quad
    \mathcal{R}^{RB-R}_{11} = 0.  \label{exuR2}
\end{align}

Thus, using Eqs.~\eqref{exuB2} and \eqref{exuR2}, the components of the energy-momentum tensor on $\Sigma_{BR}$ is given as
\begin{align}
  \sigma_{RB} &= \left. - \frac{f_R - f_B}{8 \pi r} \right|_{\Sigma_{RB}} = \frac{m(v_{V,0}) - M}{4 \pi r^2}, \label{enden1} \\
  j_{RB}^{A}&=0, \\
  P_{RB}& = 0. 
\end{align}
Therefore, the condition for the dust shell $(P=0)$ is manifestly satisfied for the junction between the black hole and the Vaidya region.

\subsubsection{Matching Vaidya and wormhole}

\begin{figure}[H]
  \centering
  \includegraphics[scale = 0.5]{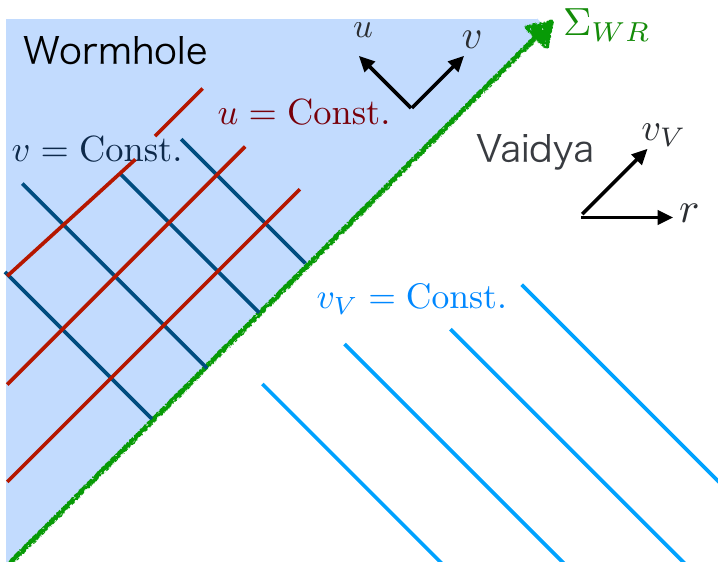}
  \caption{The picture of the right Vaidya region which is described by~\eqref{vaiRefc}, and the wormhole region which is described by~\eqref{WHuv} in the vicinity of $\Sigma_{WR}$.
  The position of the shell $\Sigma_{WR}$ is described by the equations $v_{V}-\zeta_{V}(r)=0$ and $u-v_{V,0}=0$ from the Vaidya region and the wormhole region, respectively.}
  \label{SchemWR}
\end{figure}

Next, we impose the junction to the right Vaidya region and the self-similar wormhole across $\Sigma_{WR}$ (Fig.~\ref{SchemWR}).
We shall introduce the intrinsic
coordinates on $\Sigma_{WR}$ as $\xi^{a}=(r, \theta, \phi)$.

From the right Vaidya region, $\Sigma_{WR}$ is given by the equation $\Phi^{WR-R}=v_{V}-\zeta_{V}(r)=0$, and we have
\begin{equation}
  \boldsymbol{e}^{WR}_{(1)} = \frac{2}{f_{R}}  \boldsymbol{\partial}_{v_{V}} + \boldsymbol{\partial}_{r} 
  , \quad \boldsymbol{e}^{WR}_{(2)}  = \boldsymbol{\partial}_\theta, \quad \boldsymbol{e}^{WR}_{(3)}  = \boldsymbol{\partial}_\varphi. 
\end{equation}
Here, we have normalized $ \boldsymbol{e}^{WR}_{(1)}$ so that the coefficient of $\boldsymbol{\partial}_{r}$ to be unity.
As we can see from Eq.~\eqref{norma2} below, this normalization corresponds to be taking $\alpha =1$ in Eq.~\eqref{ndPhi}.
The normal and transverse null vectors are respectively
\begin{align}
  \boldsymbol{n}^{WR} & = -  \boldsymbol{d}v_{V} + \frac{2}{f_{R}}\boldsymbol{d}r , \label{norma2}\\
  \boldsymbol{N}^{WR-R, \flat} &= - \frac{f_{R}}{2} \boldsymbol{d}v_{V}.
\end{align}

The  extrinsic curvature is
\begin{align}
  \mathcal{R}^{WR-R}_{ab} 
  & = \mathrm{diag} \qty(- \frac{2}{f_{R}^2} \pdv{f_{R}}{v_{V}}, - \frac{rf_{R}}{2}, - \frac{rf_{R}}{2} \sin^2 \theta),
\end{align}
and we obtain
  \begin{align}
    h^{AB} \mathcal{R}^{WR-R}_{AB} &= -\frac{f_{R}}{r}, \quad
    \mathcal{R}^{WR-R}_{1 B}=0 , \quad
    \mathcal{R}^{WR-R}_{1 1} = - \frac{2}{f_{R}^2} \pdv{f_{R}}{v_{V}}.  \label{exutR2}
\end{align}

From the wormhole region, $\Sigma_{WR}$ is given by the equation $\Phi^{WR-W}= u-u_{0}=0$
Using the coordinates in the wormhole region, the basis are written as
\begin{equation}
  \boldsymbol{e}^{WR}_{(1)} =\qty(\partial_{r}v) \boldsymbol{\partial}_{v}
  , \quad \boldsymbol{e}^{WR}_{(2)}  = \boldsymbol{\partial}_\theta, \quad \boldsymbol{e}^{WR}_{(3)}  = \boldsymbol{\partial}_\varphi, 
\end{equation}
and the normal can be written as
\begin{align}
  \boldsymbol{n}^{WR}  &= -\frac{1}{\alpha_{W}}\boldsymbol{d}u.
  \label{norma3}
      \end{align}
Here, we have defined
      \begin{align}
  \frac{1}{\alpha_{W}} & = \left.\frac{\mathrm{e}^{B}}{2\sqrt{A}}\partial_{r}v \right|_{u=u_{0}} \nonumber \\
  & = \left.\frac{\mathrm{e}^{B}}{A}\sqrt{\frac{v}{u}}\qty(1 + \frac{v}{u}\frac{A'(v/u)}{A})^{-1}\right|_{u=u_{0}} \\
  & = \left.\frac{2\mathrm{e}^{B + l}}{2A + A'(l)}\right|_{u=u_{0}},
    \end{align}
where in the second line, the prime denotes the derivative with respect to $v/u$ , while in the third line the prime denotes the derivative with respect to $l$.
From conditions $\boldsymbol{N} \cdot \boldsymbol{N}= 0$ and $\boldsymbol{n}^{\sharp} \cdot \boldsymbol{N} = -1$, we obtain
\begin{align}
  \boldsymbol{N}^{WR-W, \flat} &= -\alpha_{W}\frac{\mathrm{e}^{B}}{2\sqrt{A}} \boldsymbol{d}v.
    \end{align}

The transverse extrinsic curvature is
\begin{align}
  \mathcal{R}^{WR-W}_{ab} &= \mathrm{diag} \qty(\mathcal{R}^{WR-W}_{rr}, \mathcal{R}^{WR-W}_{\theta\theta}, \mathcal{R}^{WR-W}_{\theta\theta} \sin^2 \theta), \label{exu2} \\
  \mathcal{R}^{WR-W}_{rr} &= \frac{2\mathrm{e}^{-t}\sqrt{A}\qty[A'(l)\qty(B'(l) - 1) +2A(B'(l) + 1) - A''(l)]}{L \qty(A'(l) + 2A(l))^2} \nonumber \\
& = - \frac{2 A(l) \mathrm{e}^{2l}}{r \qty(A'(l)+2 A(l))^2}, \\
\mathcal{R}^{WR-W}_{\theta\theta} &= \frac{\mathrm{e}^{t-B}}{8}L \qty(4 A(l)^2 - A'(l)^2)  \nonumber \\
&= \frac{\mathrm{e}^{-B}}{8}\frac{\qty(4 A(l)^2 - A'(l)^2)}{\sqrt{A}}r, \\
\therefore h^{AB} R^{WR-W}_{AB} &= \frac{\mathrm{e}^{-B}}{4}\frac{\qty(4 A(l)^2 - A'(l)^2)}{\sqrt{A} r}  \nonumber \\
&= \frac{\qty(A'(l)^2 - 4 A^2)}{r\qty[4A^2 - 2A\cosh(2l) + A'(l)(\sinh(2l) - A'(l))]}
\end{align}
Here, we have used Eq.~\eqref{Aeq} and its derivatives, and the relation
\begin{align}
r&=\mathrm{e}^{t}L \sqrt{A(l)} \label{r} 
\end{align}
to eliminate $t$.
We note that, on $\Sigma_{WR}$, we have
\begin{align}
r = \mathrm{e}^{l}\sqrt{A(l)}u_{0}.
\end{align}
In particular, at the throat, the areal radius of the throat is given as
\begin{align}
r_{0} := r(l=0) & = \sqrt{A_{\rm{th}}}u_{0}. \label{rl0} 
\end{align}

Thus, on $\Sigma_{WR}$, we have
\begin{align}
  \sigma_{WR} &=-\frac{1}{8\pi r}\qty(\frac{A'(l)^2 - 4 A^2}{4A^2 - 2A\cosh(2l) + A'(l)(\sinh(2l) - A'(l))} + 1 -  \frac{2m(v_{V})}{r}  ), \\
  j_{WR}^{A} &= 0, \\
  P_{WR}  &= -\frac{1}{8\pi} \qty(\mathcal{R}^{WR-W}_{rr} -\frac{4r\partial_{v_{V}}}m(v_{V}){(r - 2m(v_{V}))^2}) \label{P2} \\
 &= -\frac{1}{8\pi} \qty(\mathcal{R}^{WR-W}_{rr} -\frac{2m'(r)}{r - 2 m(r)}).
\end{align}
Here, we have used Eq.~\eqref{zetaR} and the equation for the shell $v_{V}-\zeta_{V}(r)=0$.

We shall impose the condition for the vanishing pressure, $P_{WR}=0$.
 This condition can be rewritten as
\begin{align}
 \frac{2m'(r)}{r - 2 m(r)} = \mathcal{R}^{WR-W}_{rr} (r).
\end{align}
This differential equation can be solved as
\begin{align}
 m(r) &= \frac{1}{\mu(r)}\qty(m_{0} + \frac{1}{2}\int^{r}_{r_{0}} \mu \qty(r') \mathcal{R}^{WR-W}_{rr} \qty(r')r' dr'), \\
 \mu(r) &:= \exp\qty(\int^{r}_{r_{0}} \mathcal{R}^{WR-W}_{rr} \qty(r')dr').
\end{align}
Here, we have defined $m_{0} := m(r_{0})$.

It is difficult to obtain an explicit analytic expression for $m(r)$ for a generic value of $l$.
Since we aim to obtain the relation between the black hole mass and the wormhole throat radius, we focus on the region near the throat.
By introducing $\delta r:= r -r_{0}$, we have
\begin{align}
 &1 -  \frac{2m(v_{V})}{r} = 1 -  \frac{2m_{0}}{r_{0}} + O\qty(\delta r), \\
 & \frac{A'(l)^2 - 4 A^2}{4A^2 - 2A\cosh(2l) + A'(l)(\sinh(2l) - A'(l))} = \frac{2A_{\rm{th}}}{1- 2A_{\rm{th}}} + O\qty(\delta r), \\
 & \therefore \sigma_{WR} =-\frac{1}{8\pi r_{0}}\qty(1 -  \frac{2m_{0}}{r_{0}}  + \frac{2A_{\rm{th}}}{1- 2A_{\rm{th}}}) + O\qty(\delta r).
\end{align}
We see that $\sigma_{WR}<0$ for the leading-order in $\delta r$ expansion.

\subsection{Relation between mass and throat radius}

In this subsection, we shall discuss the relation between energy of the black hole and the wormhole.

First, we discuss how the mass of the initial black hole changes due to the injection of the null dust shells of mass $\Delta M$.
Here, the mass of the shell can be defined as
\begin{align}
\Delta M &:= 4 \pi r^2 \sigma_{RB} = m(v_{V, 0}) - M. \label{shemas} 
\end{align}
From this expression, we see that $\Delta M$ is constant.

Next, we shall discuss the effect of the junction of the energy-momentum tensor across the intersection of two shells.
As discussed in Ref.~\cite{Koga:2025bqw}, the condition for the junction across the intersection can be written as the continuity of the transverse vector.
Let us consider the junction of the left Vaidya region and the wormhole region across $\Sigma_{WL}$ (Fig.~\ref{Schem}).
$\Sigma_{WL}$ is represented as $u_{V} - \chi_{V}(r) =0$ in the left Vaidya region, and $v - v_{0} = 0$ (const.) in the wormhole region.
By similar computation to the junction across $\Sigma_{WR}$, the transverse vectors of the Vaidya and wormhole region are computed as  
\begin{align}
\boldsymbol{N}^{WL-L, \flat} &= - \frac{f_{L}}{2} \boldsymbol{d}u_{V}, \\
   \boldsymbol{N}^{WL-W, \flat} &= -\tilde{\alpha}_{W}\frac{\mathrm{e}^{B}}{2\sqrt{A}} \boldsymbol{d}u, \\
  \frac{1}{\tilde{\alpha}_{W}} & = \left.\frac{\mathrm{e}^{B}}{2\sqrt{A}}\partial_{r}u \right|_{v = v_{0}} \nonumber \\
  & = \left.\frac{\mathrm{e}^{B}}{A}\sqrt{\frac{u}{v}}\qty(1 - \frac{v}{u}\frac{A'(v/u)}{A})^{-1}\right|_{v = v_{0}} \\
  & = \left.\frac{2\mathrm{e}^{B+l}}{2A - A'(l)}\right|_{v = v_{0}},
\end{align}
respectively.

Since $\boldsymbol{N}^{WL-L, \flat}$ and $\boldsymbol{N}^{WL-W, \flat}$ are equal at the junction across $\Sigma_{WL}$, we have
\begin{align}
   \left.f_{L}du_{V}\right|_{\Sigma_{WL}}=\left. \frac{\tilde{\alpha}_{W} \mathrm{e}^{B}}{\sqrt{A}} du\right|_{\Sigma_{WL}}, \label{Sig1}
\end{align}
from which we can obtain the relation between $du_{V}$ and $du$ at the intersection point.

We shall compute the energy-momentum across the intersection point.
The energy-momentum tensor of $\Sigma_{LB}$ across the left Vaidya region and the wormhole region is given by
\begin{align}
   T_{LB, \mu\nu} \boldsymbol{d}x^{\mu}  \otimes  \boldsymbol{d}x^{\nu} &= \sigma_{LB}n^{LB}_{\mu}n^{LB}_{\nu}\delta(u_{V} - u_{V, 0})\boldsymbol{d}x^{\mu}  \otimes  \boldsymbol{d}x^{\nu} \nonumber \\
     &= \frac{\sigma_{LB}}{2f_{L}} \times f_{L}\qty(\boldsymbol{d}u_{V}  \otimes  \boldsymbol{d}
     \Theta(u_{V} - u_{V, 0}) +  \boldsymbol{d}
     \Theta(u_{V} - u_{V, 0}) \otimes \boldsymbol{d}u_{V} 
     ).
\end{align}
Here, we have used the relation $\boldsymbol{n}^{LB} = -\boldsymbol{d}u_{V}$, which can be obtained similarly  as $\boldsymbol{n}^{RB}$
For the junction at $\Sigma_{WR}$, the energy-momentum tensor is
\begin{align}
   T_{WR,\mu\nu}\boldsymbol{d}x^{\mu}  \otimes  \boldsymbol{d}x^{\nu} &= \alpha_{W}\sigma_{WR}n^{WR}_{\mu}n^{WR}_{\nu}\delta(u -u_{0})\boldsymbol{d}x^{\mu}  \otimes  \boldsymbol{d}x^{\nu}\nonumber \\
     &= \frac{\sigma_{WR}}{2\alpha_{W}\tilde{\alpha}_{W}}\frac{\sqrt{A}}{\mathrm{e}^{B}} \times \frac{\tilde{\alpha}_{W}\mathrm{e}^{B}}{\sqrt{A}}
     \qty(\boldsymbol{d}u \otimes \boldsymbol{d}\Theta(u -u_{0}) + \boldsymbol{d}\Theta(u -u_{0}) \otimes \boldsymbol{d}u).
\end{align}

We shall require continuity of the surface energy-momentum tensor across the crossing point.
Using Eq.~\eqref{Sig1} and left-right symmetry of the spacetime, this condition can be written as
\begin{align}
   \lim_{l\rightarrow 0}\frac{\sigma_{LB}}{f_{L}} &= \lim_{l\rightarrow 0}\frac{\sigma_{WR}}{\alpha_{W}\tilde{\alpha}_{W}}\frac{\sqrt{A}}{\mathrm{e}^{B}}. \label{sigeq}
   \end{align}
The left hand side and right hand side of Eq.~\eqref{sigeq} is respectively expressed as
   \begin{align}
 \lim_{l\rightarrow 0}\frac{\sigma_{LB}}{f_{L}} &=  \frac{\Delta M}{4\pi r_{0}^2 f_{L}} , \\
 \lim_{l\rightarrow 0}\frac{\sigma_{WR}}{\alpha_{W}\tilde{\alpha}_{W}}\frac{\sqrt{A}}{\mathrm{e}^{B}} &\simeq  \sigma_{WR}\qty(\frac{1}{2A_{\rm{th}}}-1).
\end{align}
Here, we have approximated as $\alpha_{W}  \simeq A(l)\mathrm{e}^{-B-l}$ and $\tilde{\alpha}_{W} \simeq A(l)\mathrm{e}^{-B+l}$.
Thus, we obtain
\begin{align}
\Delta M = - \frac{r_{0}}{2} \left( 1 - \frac{2 m_{0}}{r_{0}}\right)
\left[ 1 + \left(1 - \frac{2m_{0}}{r_{0}}\right) \frac{1 - 2 A_{\text{th}}}{2 A_{\text{th}}} \right].
\label{DM}
\end{align}
From this relation, we can see that the  change in the mass from the black hole to the Vaidya region is negative.
This is compatible with our expectation that we need to add a negative energy for the wormhole formation from a black hole.

Furthermore, by substituting Eq.~\eqref{DM} into Eq.~\eqref{shemas} with $m(v_{V, 0}) = m_{0}$, we may solve $A_{\rm{th}}$ in terms of $M$, $\Delta M$ and $r_{0}$ as
\begin{align}
A_{\text{th}} = \frac{1}{2} \frac{\left(1 - \frac{2 (M + \Delta M)}{r_{0}}\right)^2}{  \left(1 - \frac{2 (M + \Delta M)}{r_{0}}\right)^2 - \left(1 - \frac{2M}{r_{0}} \right)} \label{Ath}
\end{align}

The condition $A_{\rm{th}} <1/2$ implies $r_{0} < 2 M$, which means that in the Schwarzschild black hole, the location $r_{0}$ that will become the wormhole throat lies inside the event horizon.
This is consistent with our dynamical formation scenario: the negative-energy flux reduces the black hole mass, causing the horizon to shrink and eventually exposing the throat region that was previously hidden behind it.
Thus, the relation~\eqref{Ath} indicates that the wormhole throat emerges from within the original black hole interior.
We also find that $\partial A_{\rm{th}} /\partial(\Delta M)<0$ for $r_{0} < 2 M$.
This means that $A_{\rm{th}}$ is larger for the larger (negative) change of the mass.

Here, we emphasize a key difference from the static cases~\cite{Hayward:2004wm,Koyama:2004uh,Koga:2025bqw}: in our case, the throat radius depends not only on the black hole mass but also on the parameter $u_{0}$, which specifies the insertion time of the shell.
This dependence directly reflects the fact that the wormhole throat expands as time progresses.

\section{Conclusion and discussion}
\label{Section}

In this work, we explored a non-static extension of the Hayward-Koyama wormhole formation scenario~\cite{Hayward:2004wm,Koyama:2004uh}.
Our construction employed colliding streams of negative-energy null dust as the matter source and imposed self-similar symmetry to obtain a tractable class of solutions.
This approach allowed us to construct wormhole spacetimes with regular throats and embed them consistently into a dynamical model describing the transition from a black hole to a wormhole.
Under this symmetry, we solved the equations of motion numerically. Boundary conditions were imposed to enforce symmetry under inversion of the radial coordinate and to ensure that the areal radius attains a minimum at the symmetric point.

We found that the large-$|l|$ asymptotics of the solutions are governed by a single parameter, $\gamma$.
For $\gamma > 2$, pp-curvature singularity develops in the future null direction, associated with the divergence of the negative-energy flux (Eq.~\eqref{Tppas}).
When $\gamma > 4$, additional scalar curvature singularities arise in the future null direction as well as in the spatial and past null regions, as seen from the expressions~\eqref{RRas} and \eqref{CCas}.
By contrast, for $\gamma < 2$, the solutions avoid singularities at both spatial and future null infinity, while still retaining an unavoidable singularity in the past direction.
Moreover, larger throat radii naturally correspond to smaller values of $\gamma$ (Fig.~\ref{fig:gamma}), ensuring that spacetimes with sufficiently large throats approach future infinity smoothly.
Thus, the throat size serves as the key control parameter that determines whether the wormhole geometry remains regular or develops singularities.

These findings reveal a novel mechanism by which sufficiently large wormhole throats can evade singularities in the spatial and future null directions, in sharp contrast to previous static constructions.
At the same time, the persistence of an initial singularity underscores the limitations of self-similarity: the wormhole effectively emerges from an initial singularity.

Building on this, we presented a global formation model, schematically depicted in Fig.~\ref{Schem}.
By patching together Schwarzschild, Vaidya, and self-similar wormhole geometries across thin null shells, we constructed a consistent scenario in which an initial black hole is converted into a traversable wormhole via the injection of negative-energy flux.
The Barrabès–Israel formalism~\cite{Barrabes:1991ng} provided the junction conditions, and their explicit evaluation showed how the throat radius is determined by the integrated mass loss during the Vaidya phase (Eq.~\eqref{Ath}).
In particular, the vanishing-pressure condition on the shells yields a first-order ordinary differential equation for the Vaidya mass function, fully fixing the flux profile in terms of the wormhole data.
This establishes a concrete mechanism through which negative-energy radiation can dynamically generate a wormhole.
Moreover, when the throat size is sufficiently large, this process can be interpreted as dynamically resolving the initial black hole singularity through a thin shell of negative-energy null dust.

Several open issues remain. First, the physical mechanism underlying singularity avoidance for large throats deserves further investigation.
In particular, clarifying the role of negative-energy fluxes in shaping the asymptotic structure could provide crucial insights.
Second, the perturbative stability of these self-similar wormholes must be addressed to assess their viability as time-dependent solutions.
Finally, extensions beyond spherical symmetry or the inclusion of additional matter fields may uncover richer classes of singularity-avoiding wormholes.

\section*{Acknowledgments}

This work was partially supported by Grants-in-Aid for Scientific Research from the Japan Society for the Promotion of Science (JSPS) and the Ministry of Education, Culture, Sports, Science and Technology (MEXT) of Japan
under Grant Numbers
JP23H01170~(YK),
JP24KJ1223~(DS), 
JP21H05189~(DY).

\appendix

\section{Definitions of non-static wormholes}
\label{throat}

In non-static settings several inequivalent definitions of wormhole throats have been discussed in the literature. 
In this appendix, let us review the definition by Maeda, Harada, and Carr~\cite{Maeda:2009tk}, which proposed a definition suitable for cosmological wormhole solutions, and apply to our problem. 
Their motivation was that in spacetimes with an initial singularity there is, in general, no past null infinity and no trapping horizon. 
As a result, earlier definitions based on null expansions, such as those by Hochberg–Visser~\cite{Hochberg:1997wp} and Hayward~\cite{Hayward:1998pp}, fail to capture physically interesting wormhole geometries in cosmological scenarios.

The key idea of the definition in Ref.~\cite{Maeda:2009tk} is to identify the throat on a spacelike hypersurface as a two-sphere of nonvanishing minimal area.
Concretely, introducing a radial spacelike vector 
$\boldsymbol{s}$, the conditions are
\begin{align}
 \bm{\nabla}_{\bm{s}}  R &= 0, \\
\bm{\nabla}_{\bm{s}} \bm{\nabla}_{\bm{s}}  R &> 0 ,
\end{align}
where $R$ is the areal radius.
This criterion generalizes the static minimal-surface condition to time-dependent spacetimes, while avoiding the restriction to null hypersurfaces. A throat defined may coincide with a bifurcating trapping horizon (the “momentarily static” case), but more generally it can lie inside a trapped region without being associated with any trapping horizon.
In this sense, this definition encompasses 
cosmological wormholes that would otherwise be excluded from other frameworks such as Refs.~~\cite{Hochberg:1997wp,Hayward:1998pp}

We now apply this definition to our self-similar solutions. 
For a spherically symmetric spacetime, it is natural to take $\boldsymbol{s}=\boldsymbol{\partial}_{l}$.
Then we obtain
\begin{align}
\bm{\nabla}_{\bm{s}}  
R&=\frac{\mathrm{e}^
{t}LA'(l)}{2\sqrt{A}}, \\
\bm{\nabla}_{\bm{s}}  \bm{\nabla}_{\bm{s}}  R &=\frac{\mathrm{e}^
{t}L}{8A^{3/2}}\qty(4A\cosh(2l) - A'(l)\sinh(2l)).
\end{align}
Under the boundary conditions~\eqref{BC}, these reduce in the 
candidate throat $l=0$ to
\begin{align}
\bm{\nabla}_{\bm{s}} R&=0, \\
\bm{\nabla}_{\bm{s}}  \bm{\nabla}_{\bm{s}} R &=\frac{\mathrm{e}^
{t}L}{2A_{\rm{th}}}>0.
\end{align}
showing that the throat is located at $l=0$ irrespective of the 
value of $t$. 
Thus, the definition in Ref.~\cite{Maeda:2009tk} consistently identifies the wormhole throat in the solutions considered in this paper, and it is the definition we adopt throughout.

\bibliography{bibs/ref}
\bibliographystyle{JHEP.bst}

\end{document}